%% file: article_sigma_arxiv.tex
\documentclass[preprint,5p,twocolumn]{elsarticle}
\usepackage[latin1]{inputenc}  
\usepackage[T1]{fontenc}       

\usepackage{graphicx}

\usepackage{amssymb}

\biboptions{sort&compress}

\journal{\ldots}

\begin{document}

\begin{frontmatter}

\title{Dispersion of longitudinal momentum distributions induced in fragmentation reactions}

\author[a,b]{A. Bacquias\corref{cor1}\fnref{1}}
\ead{bacquias@gmail.com}
\cortext[cor1]{corresponding author}
\author[a,c]{V. Föhr}
\author[a]{D. Henzlova\fnref{2}}
\author[a]{A. Keli\'{c}-Heil}
\author[a]{M.V. Ricciardi}
\author[a]{K.-H. Schmidt}
\address[a]{GSI Helmholtzzentrum für Schwerionenforschung GmbH, Darmstadt, Germany}
\address[b]{Université de Strasbourg, France}
\address[c]{Jyväskylän Yliopisto, Jyväskylä, Suomi/Finland}
\fntext[1]{Present address: CNRS-IPHC Strasbourg, France}
\fntext[2]{Present address: Los Alamos National Laboratory, NM, USA}

\begin{abstract}
 On the basis of systematic measurements of fragmentation reactions, which provide a detailed overview on the velocity distributions of residual nuclei, an improved description of the kinematical properties of the fragmentation residues is established. 
This work is dedicated to the fluctuations of their momentum distributions. 
In contrast to previous investigations, limited to close-to-projectile fragments, we extended our study to the entire production range, down to the lightest observed fragments. 
In this context, beside the contribution of abrasion and evaporation processes, we considered the effect of the thermal break-up on the width of the momentum distributions. Using approximated theoretical descriptions of the different reaction stages, a new analytical formula for the variance of the momentum distribution is derived, which is well adapted to technical applications. 
\end{abstract}


\end{frontmatter}


\input{introduction.tex}

\input{data.tex}

\input{fragmentation.tex}

\input{model.tex}

\input{conclusion.tex}

\appendix
\input{appendix.tex}

\bibliographystyle{model1-num-names}
\bibliography{biblio}

\end{document}

%% file: introduction.tex
\section{Introduction}
Heavy-ions reactions with various target materials at energies well above the Fermi-energy regime, have gained increasing interest for a variety of applications. They are used for the production of rare isotopes in existing~\cite{iso,gsi,msu,riken} and future secondary-beam facilities~\cite{fair,ria,eurisol}, they are exploited in the field of nuclear technology for the operation of spallation neutron sources~\cite{sns,wag}, in particular for the incineration of nuclear waste in an accelerator-driven system~\cite{alander}, they are a source of activation due to inevitable beam losses along high-energy accelerators and in the beam dump~\cite{ulrici}. Besides their cross sections, the kinematical characteristics of the fragmentation residues emerging from these reactions are very important. In in-flight secondary-beam facilities, they determine the emittance of the rare-isotope beams; in irradiated construction material, they are responsible for aging phenomena. 

In non-central collisions of heavy ions at energies far beyond the Fermi-energy regime, the geometrical abrasion model~\cite{bow} suggests a clear cut between participant matter, corresponding to the overlap zone of the two nuclei, and spectator matter outside the overlap zone. While the participant matter is subject to strong heating and compression due to high-energy nucleon-nucleon collisions, the spectator matter is sheared off from the projectile, respectively target nucleus, and continues moving on with essentially its initial velocity. On a smaller scale, however, a shift in the mean longitudinal velocity of the spectator matter is observed. For very peripheral collisions, friction dominates~\cite{fric}, while for less peripheral collisions an opposite effect has been observed~\cite{ric}, which has been attributed to the response of the spectator to the participant blast~\cite{shi}. In addition, the velocity distribution of the spectator matter is characterized by fluctuations around its mean value. The magnitude of these fluctuations is most often larger than the shift of the mean value and, thus, constitutes the most important characteristics of the kinematical properties of the spectators. 

The basic theoretical understanding of the fluctuations of the momentum distributions dates back to the pioneering work of Goldhaber in 1974~\cite{gol}, who described two possible scenarios, the resulting recoil momentum due to a sequential particle emission, or the random superposition of the individual Fermi momenta of the nucleons removed in an abrasion process. 
A quantitative simulation of the velocity distributions of individual nuclides, also concentrating on very peripheral collisions, considering the contributions of the abrasion and the evaporation stages by a full Monte-Carlo nuclear-reaction code, has been presented by Hanelt {\it et al.}~\cite{han}. They have shown that the recoil momentum of evaporated particles modifies the momentum dispersion of final fragments. 
A comprehensive systematics on momentum distributions from reactions with relativistic ions, documenting the status of that time, was established by Morrissey in 1989~\cite{mor}, and an empirical formula for very peripheral collisions was proposed: 
\begin{equation} 
\sigma^2_{p_\parallel}=\frac{150^2}{3} \cdot (A_p-A_{f})
\label{morissey}
\end{equation}
where $A_p$ and $A_f$ represent the mass of the projectile and the mass of the final fragment, respectively. 
Although this empirical formula perfectly reproduces most data near the
projectile, it becomes unrealistic for lighter fragments. Indeed, the predicted dispersion prediction does not decrease towards low masses and finally gives unrealistically large values for very light fragments. One has to realize that  the width of the momentum has to go down to zero with mass approaching zero!
Thus, this formula cannot cover the whole range of fragments produced and observed in fragmentation reactions. 

Recently, a semi-empirical model describing the momentum distribution of the fragmentation residues has been introduced by Tarasov~\cite{tarasov}. The momentum distributions is represented by a convolution of a Gaussian with an exponential tail, where the shape of the convoluted distribution and its variation with projectile energy and the masses of projectile and fragment are associated with the influence of abrasion, friction and evaporation on the kinematics of the fragments. The free parameters of that distribution are tuned to fit experimental data. 

The need to build up a model fully based on theoretical considerations, taking into account the multiple processes at play in fragmentation reactions, valid over the whole mass range of produced residues but compact and fast enough to be used for various applications, is then obvious. 

Thanks to a series of extensive experiments on systematic measurements of fragmentation residues, mostly performed at GSI, Darmstadt, the empirical knowledge in this field has improved substantially~\cite{han,greiner,westfall,hubele,therm,weber-exp,magel,universality,dejong-exp,junghans,benlliure-u,reinhold,fritz,benlliure-npa,benlliure-npa-err,timo-u,ala,timo-leadH,fanny,transverse,freeze,timo-leadD,stolz,indra,taieb,glad,armbruster-exp,napolitani04,ric-npa,daniela-npa,ric06,audouin,enrique-D,kurcewicz,napolitani07,villagrasa,trautmann:064606,daniela-prc,benlliure-xe,alvarez-epja,alvarez-prc,daniela-JPG}. By investigating the projectile-like fragments, a detailed overview on cross sections and velocity distributions of individual nuclides has been obtained. The high quality of these data comes not only from the high precision, but also from the fact that, in contrast to experiments on the properties of target-like fragments, which suffer from a low-energy cut-off, the kinematical properties of the projectile-like fragments can be fully measured. In addition, for a few reactions, the final residues were measured over a very broad mass range. The availability of new, high-quality data has motivated us to re-visit the understanding of the kinematical properties of the fragmentation residues. The present work is dedicated to the fluctuations of their momentum distributions. 

It is our aim to study the fluctuations of the momentum distributions of all fragmentation residues emerging from the spectator matter. In contrast to many previous investigations, we consider the whole mass range, from the projectile, respectively target nucleus, down to light charged particles. 

For this purpose, we also treat the influence of the multifragmentation stage on the kinematical properties of fragmentation residues. This break-up of the system due to thermal instabilities becomes important at high beam energies and was not considered in previous works. 

In the present paper, we develop a model fully based on theory. We derive  analytical approximations from appropriate theoretical descriptions of the different stages of the collision. 
This is used to develop a new comprehensive analytical formula for the variance of the momentum distributions, well adapted to technical applications. 


%% file: data.tex
\section{Experimental data}
From the body of data on kinematical properties of fragmentation residues, we have chosen three experiments performed at the FRagment Separator (FRS) in GSI, Darmstadt. They represent the fragmentation of two light nuclei ($^{56}$Fe~\cite{villagrasa}, $^{86}$Kr~\cite{weber-exp}) and a medium-heavy nucleus  ($^{136}$Xe \cite{daniela-prc}). These experiments cover a broad mass range: residues have been observed from the vicinity of the projectile down to elements lighter than neon. 

The FRS is characterized by a high momentum resolution and a limited acceptance both in momentum and angle \cite{FRS}. The momentum resolution (FWHM) amounts to $\Delta p/p \simeq 5\cdot 10^{-4}$, the momentum acceptance is $\pm 1.5\% $, while the angular acceptance comprises $15$ mrad around the beam axis. The limited momentum acceptance is not crucial, since the full momentum range can be covered by combining results obtained for different settings of the magnetic fields. We limit the present work to the longitudinal momentum distribution of the residues emitted within the angular acceptance of the FRS. 
Nevertheless, the same formalism can be used for the transverse component of momentum distributions \cite{greiner,feshbach-huang}. Since at the projectile energies considered in the present work, the production of fragmentation residues is approximately isotropic \cite{greiner,feshbach-huang,gol-annrev}, our description should be valid for the transverse momentum dispersion as well. 

\begin{figure}[!ht]
\centering
\includegraphics*[width=85mm,height=110mm]{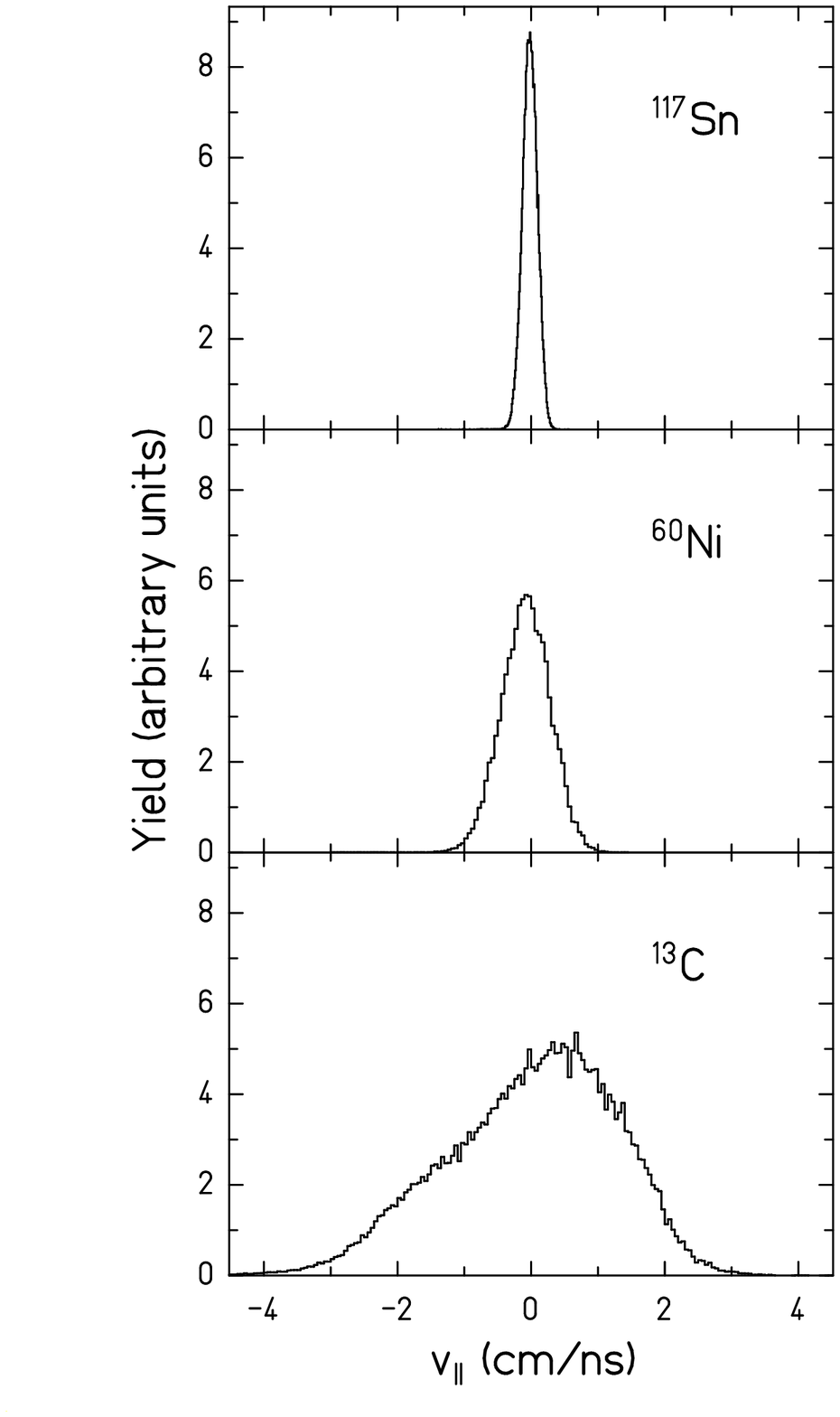}
\caption[]{\label{gaussians}\footnotesize{Longitudinal velocity distributions in the projectile frame for various fragments (transmitted through the 15~mrad angular acceptance of the FRS) from the reaction $^{136}$Xe+Pb at 1~GeV per nucleon, taken from \cite{daniela-prc}. Heavy fragments follow a Gaussian distribution. On the contrary, the lightest ones populate an asymmetric  distribution. }}
\end{figure}

Three measured longitudinal velocity distributions, in the projectile frame, are
given as examples (fig. \ref{gaussians}). They correspond to $^{117}$Sn, $^{60}$Ni and $^{13}$C fragments from the reaction $^{136}$Xe+Pb at 1 $A$ GeV \cite{daniela-prc}. Over a large mass range, the residues follow a Gaussian distribution. Only the lightest residues show a more complex distribution with an asymmetry or even several humps. 
Light nuclei can originate from two different processes \cite{ric06,napolitani07,bacquias,napolitani-jpg}. In binary mass splits, e.g. very asymmetric fission, Coulomb forces lead to a two-humped longitudinal velocity distribution \cite{timo-u,ric06}. In case of multifragmentation, the Gaussian shape of longitudinal velocity spectra is preserved. Please note that our model aims at describing fragmentation reactions, and therefore cannot be used to calculate the momentum width of fission residues. 

The data from ref. \cite{weber-exp,villagrasa,daniela-prc} will be compared with the predictions of our analytical model in the section \ref{model}, but let us first present its fundamental ideas.

%% file: fragmentation.tex
\section{Fragmentation mechanism}
Let us remind the different descriptions of the processes that happen during the collision and the de-excitation phase of the fragmentation reaction and study their influence on the width of the longitudinal momentum distributions of the fragments. 

As a typical collision configuration, we will assume a projectile nucleus hitting a target nucleus (at rest) with an impact parameter in principle different from zero. In the overlap region, projectile and target nucleons interact strongly, while the rest of the projectile and the target nucleus, so called spectators, are almost undisturbed by the collision. This picture is well suited in a wide energy domain above the Fermi energy, and the model developed in the present work can thus be used for comparison with various projectile- and target-fragmentation data from available devices (e.g. at GSI, MSU, RIKEN\ldots). In the following, we will concentrate on the projectile-like fragments, since we will compare the predictions of the present model with the projectile-like residues measured at the FRS in GSI. Of course, the model is also applicable to target-like residues. We will discuss the  contributions of the different stages of the fragmentation process to the momentum dispersion of the final residues.

%

\subsection{Abrasion}

Dedicated experiments have long shown that the longitudinal momentum of the heavy fragmentation residues follows a Gaussian distribution in the projectile frame. 
A few models have been developed to describe the standard deviation of this distribution. 

For the standard deviation of momentum in longitudinal or transverse directions ($\sigma_{p_\parallel}$ and $\sigma_{p_T}$), the most often used theoretical model is Goldhaber's prediction for the dispersion induced in the abrasion process due to the Fermi motion of the nucleons~\cite{gol}. 

A proposition of Goldhaber is to consider the reaction as a sudden cut-off of a part of the projectile, without taking into account any further evolution of the remaining part of the projectile. 
Some nucleons are removed instantaneously, without inducing momentum transfer. This reaction mechanism is referred to as abrasion, and it suggests already a removal of matter by friction phenomena, so that its description in~\cite{gol} is probably too simple, but we shall discuss that after considering the strong implications of the assumption made by Goldhaber. 
An instantaneous removal of several nucleons from the projectile does indeed affect the dynamical features of this projectile. In fact, this contribution explains great part of the measured momentum width of the fragmentation residues. The removed nucleons are well defined by their positions inside the nucleus at the moment of collision. Yet, the momenta of these nucleons are sampled over a broad distribution. 

Considering the ensemble of nucleons as a Fermi gas, we know that these constituents of the nucleus have an intrinsic movement, even at zero temperature. This internal motion affects the observable features of the projectile-like fragments. 
In a Fermi gas inside a square-well potential with infinitely high borders, there is no correlation between the position and the momentum of a nucleon. 
Thus, the momentum of an abraded nucleon is a random sample from the projectile's Fermi sphere filled up to the radius $p_F$, in Goldhaber's view~\cite{gol}. The authors of the present paper are aware of the works of Friedman~\cite{friedman-prc27} demonstrating that absorption, in peripheral collisions, prevents from sampling over the whole nucleus. Nevertheless, as one can see from ref.~\cite{friedman-prc27}, the predictions with the absorption taken into account are rather close to the Goldhaber model. Therefore, and owing  to the transparency of that latter, it was decided to use Goldhaber model as a basis of the present work. 

The mean square momentum of a nucleon in a Fermi gas is 
\begin{equation}
<p^2_{\mathrm{fermion}}> = \frac{\int _0 ^{p_F} p^2 \: \mathrm{d}^3p}{\int _0 ^{p_F} \: \mathrm{d}^3p} = \frac{3}{5}p^2_F \mbox{ .}
\label{p_Fermi}
\end{equation}
The momentum variance $\sigma_p$ of the projectile spectator is given by the sum of the individual contributions from the abraded nucleons. 
Provided the assumption that this broadening is equally distributed in all directions, the projection along the beam axis gives $\sigma^2_{p_\parallel}=\sigma^2_p/3$ (it is also relevant for transverse momentum studies, see for example~\cite{transverse}). 

The variance is then linked with the Fermi momentum $p_F$ of the projectile of mass $A_p$ and the mass of the fragment after the abrasion process, so-called prefragment, $A$:  
\begin{equation} 
\sigma^2_{p_\parallel}=\frac{p_F^2}{5} \cdot \frac{A(A_p-A)}{(A_p-1)}\mbox{ .}
\label{goldhaber}
\end{equation}
The Fermi momentum varies only weakly with the mass of the nucleus. 
One can find an expression of the mass dependence of the Fermi momentum in reference~\cite{giacomelli} that relies on data from~\cite{moniz}: 
\begin{equation}
p_F(A) = 281 \cdot (1-A^{-0.568}) \: \mathrm{MeV}/c \mbox{ .}
\label{fermi}
\end{equation}
We can comment that eq. \ref{goldhaber} is symmetric regarding the mass, so that an abrasion process that removes half of the nucleons in the projectile produces the largest width in the distribution. In this hypothesis, light fragments and fragments of mass close to that of the projectile should have a similar broadening of their longitudinal momentum distributions. 

This model reproduces some data rather well. At the time of release of Goldhaber's article, most data available were limited to small mass losses, corresponding to the heaviest fragments. Goldhaber's formula is indeed in quite good agreement with these data. 

But Goldhaber's model is incomplete in the sense that the spectator matter after the abrasion process is excited. Therefore, this model does not refer to the measured fragmentation residues, which result from a subsequent deexcitation  stage, but rather to an excited prefragment. The comprehension of the deexcitation processes is the key to address a model for the final fragment. 

\subsection{Evaporation}
\label{evaporation}

The removal of nucleons by abrasion is a step that does not leave the nucleus cold. 
It can be seen as a friction process, removing nucleons from the projectile, but also inducing excitation energy in the remaining fragment. 
It was shown by different methods~\cite{therm,harvey} that this energy induction amounts on average to 27~MeV per abraded nucleon. 
Therefore, the spectators may be highly excited, giving rise to the evaporation of a considerable number of nucleons and light nuclei. 

The evaporation stage can be described by an appropriate code in the frame of the statistical abrasion-ablation model~\cite{abrabla}, using realistic binding energies, Coulomb barriers and level densities. Each emitted particle induces a recoil momentum to the corresponding residue according to its kinetic energy and momentum conservation. 
The influence of the evaporation process has been demonstrated by Hanelt {\it et al.}~\cite{han}, stating that the recoil-momentum induced by sequentially evaporated nucleons should noticeably modify the width of momentum distributions. 

In most cases, the individual contributions of the evaporated particles are small and just slightly increase the width of the Gaussian distribution resulting from the abrasion stage. Only in specific cases, like symmetric fission, the recoil momentum is so large that it dominates the kinematic properties of the final fragments, and deviations from a Gaussian distribution are observed. Although  fission plays an important role for very heavy fragments, we will consider only non-fissioning systems in the present work. 

De Jong {\it et al.}~\cite{dej} included the influence of the evaporation process on the {\em angular} momentum distribution in an analytical formula. The basic idea was to assume a certain proportionality between abraded mass and evaporated mass. They obtained satisfactory predictions with a single parameter. They used $\overline{\nu}=2$ as a typical value of mass loss due to evaporation per abraded nucleon. This means that the abrasion of one nucleon leads on average to a mass loss of two nucleons by evaporation. Their description was formulated as: 
\begin{equation} 
\sigma^2_J=0.16 A^{2/3}_p \cdot \frac{(\overline{\nu} A_p+A)(A_p-A)}{(\overline{\nu}+1)^2(A_p-1)} \mbox{ .}
\label{dejong}
\end{equation}

We have adapted this idea for expressing the consequence of evaporation on the {\em linear} momentum by a dedicated analytical expression. It will be presented in section \ref{model}. 


For the prefragment, the release of nucleons (mostly individually but also as bound clusters) decreases its excitation energy. 
Taking into account the mean separation energy of one nucleon and the average
nuclear temperature inside the prefragment, this ``cooling'' of the system allowed by evaporation is considered to be 15~MeV per emitted nucleon on average. 

Since the angular momentum induced in an abrasion reaction is rather low~\cite{dej}, evaporation from the excited pre-fragment is an isotropic process in the frame of the source of emission. 
Thus, the mean velocity of the prefragment is not affected by evaporation. The momentum, however, is reduced on average because of the mass loss. In addition, the recoil momenta of the individual evaporated particles tend to increase the width of the momentum distribution.  

In order to compare explicitly the impact of the recoil process with the Fermi contribution discussed before, let us define $\eta^2$, the ratio between the mean individual energy taken from the prefragment by the evaporation of one nucleon $<E_{\mathrm{evap}}>$ and the mean individual energy of one nucleon inside the nucleus due to the Fermi motion $<E_{\mathrm{fermion}}>$: 
\begin{equation}
\eta^2=\frac{<E_{\mathrm{evap}}>}{<E_{\mathrm{fermion}}>} \mbox{ .}
\end{equation}

The mean kinetic energy $<E_{\mathrm{fermion}}>$ of a nucleon in a Fermi gas is:
\begin{equation}
<E_{\mathrm{fermion}}> = \frac{\int _0 ^{p_F} E \: \mathrm{d}^3p}{\int _0 ^{p_F} \: \mathrm{d}^3p} = \frac{3}{5}E_F \mbox{ .}
\label{E_Fermi}
\end{equation}
The average Fermi energy $E_F$ of a nucleus amounts to about 33~MeV; this value leads to $<E_{\mathrm{fermion}}>\simeq 20$MeV. 

The value of $<E_{\mathrm{evap}}>$ is not easily estimated, because it depends on the charge of the mother nucleus, and of course is different for each type of evaporated particle (neutron, proton, alpha, intermediate-mass fragments...). 
For simplicity, we consider only the evaporation of neutrons and protons. Moreover, we assume that they are emitted with equal probability, so that the mean recoil energy $<E_{\mathrm{evap}}>$ is given by the mean value between neutron and proton kinetic energies $E_n$ and $E_p$, respectively. We fix the thermal contribution to $E_n$ and $E_p$ to 8~MeV; for protons, we estimate the contribution from Coulomb repulsion using the mass $A_p$ and nuclear charge $Z_p$ of the projectile. In this way, we obtain for $\eta^2$ the following expression: 
\begin{equation}
\eta^2 = \frac{1}{2}\cdot\left[ 8 + \left( 8+ \frac{Z_p e^2}{\mathrm{r_0}(A_p^{1/3}+1)} \right) \right] / <E_{\mathrm{fermion}}> \mbox{, }
\label{eta}
\end{equation}
with $\mathrm{r_0}$=1.4~fm and $e$ the elementary charge. 

The mean square momenta of a single nucleon, for either Fermi-gas contribution or for the evaporation recoil, are directly proportional to the corresponding mean energy: 
\begin{equation}
<p^2> = 2 \mathrm{m_n} \cdot <E>
\label{E-p}
\end{equation}
with $\mathrm{m_n}$ the mass of a nucleon. 
The ratio between the mean energies is then equivalent to the ratio of momenta squared: 
\begin{equation}
\eta^2=\frac{<E_{\mathrm{evap}}>}{<E_{\mathrm{fermion}}>} = \frac{<p^2_{\mathrm{evap}}>}{<p^2_{\mathrm{fermion}}>} = \frac{5}{3} \cdot \frac{<p^2_{\mathrm{evap}}>}{p^2_F} \mbox{, }
\label{ratii}
\end{equation}
using the expression of $<p^2_{\mathrm{fermion}}>$ from eq. \ref{p_Fermi}. 
For the mean square momentum in the longitudinal direction, one finally has: 
\begin{equation}
<p^2_{\parallel \mathrm{evap}}> = \eta^2 \cdot \frac{p^2_F}{5} \mbox{ .}
\label{ratio}
\end{equation}
We shall use this expression to take the recoil momentum into account in the formula for $\sigma^2_{p_\parallel}$ (see details in section ``analytical model''). 

\subsection{Multifragmentation}

It is commonly accepted that a nucleus starts to show some thermal instabilities when its excitation energy exceeds a value of about 3~MeV per nucleon. This value can easily be reached through the abrasion process. The kinematics of the fragments is defined when they cease to interact, at freeze-out. 
The momenta of the fragments are given by a radial expansion, a random motion of the nascent fragments and by a consecutive Coulomb expansion. 
\subsubsection{Random motion}
For estimating the random motion, two rather diverging ideas were introduced in the literature. In one extreme, it was assumed that the fragments behave like molecules of an ideal gas~\cite{smm}. In this case, their individual motion is governed by a Maxwell-Boltzmann distribution and the variance of the momentum distribution of a fragment of mass $A$ produced in the break-up of a system of mass $A_{\mathrm{bu}}$ by multifragmentation is given by the following formula~\cite{gol}: 

\begin{equation} 
\sigma^2_p=\mathrm{m_n k_B} T_{\mathrm{bu}} \cdot \frac{A(A_{\mathrm{bu}}-A)}{A_{\mathrm{bu}}}
\label{goldhaber2}
\end{equation}
where $T_{\mathrm{bu}}$ is the freeze-out temperature, $\mathrm{k_B}$ is the Boltzmann constant and $\mathrm{m_n}$ denotes the nucleon mass. 

In the other extreme, the nascent fragments keep the Fermi motion of the individual nucleons in the common source~\cite{ala}. In this second case, the distribution follows the same functional form, but the temperature parameter cannot be interpreted as a real temperature. 

Odeh {\it et al.} have shown that the apparent temperature $T_{\mathrm{app}}$ extracted from the kinetic-energy spectra of the multifragmentation products is much larger than the temperature extracted using isotopic or excited-states population thermometer~\cite{ala}. Such large apparent temperatures would indicate that the process of simultaneous break-up is rather fast and that the system does not have time to reach full thermal equilibrium. In other words, the Fermi motion inside the multifragmenting source is mostly responsible for the broadening of the momentum distribution of the multifragmentation products. Thus, ideas behind the Goldhaber formula (eq.~\ref{goldhaber}) should also be valid for describing the momentum dispersion of the multifragmentation products. However, one should not forget that in case of multifragmentation the multifragmenting source is not cold as in case of abrasion, but has a finite temperature $T_{\mathrm{bu}}$. Thus, the mean velocity of the nucleons inside the multifragmenting source is larger than the mean velocity of the nucleons inside the cold source. Consequently, the  slope parameter (i.e. apparent temperature) of the kinetic-energy spectra of the multifragmentation products is increased relative to the case of the cold fragmenting source.  This effect has been studied by W. Bauer~\cite{bauer}, and he developed an analytical expression for calculating the momentum dispersion of a multifragmenting Fermi gas at finite temperature. According to ref.~\cite{bauer}, the apparent temperature $T_{\mathrm{app}}$ can be expressed as a function of the real source temperature $T_{\mathrm{bu}}$ through a Taylor expansion by: 
\begin{equation}
T_{\mathrm{app}} \approx \frac{A_{\mathrm{bu}}-A}{A_{\mathrm{bu}}-1} \frac{2}{5}E_F\! \left[ 1+ \frac{5 \pi^2}{12} \left( \frac{T_{\mathrm{bu}}}{E_F}\right)^2 \!\!\! + \mathcal{O}\! \left( \frac{T_{\mathrm{bu}}}{E_F}\right)^4 \right] \mbox{,}
\label{temp_bauer}
\end{equation} 
with $A_{\mathrm{bu}}$ the mass of the multifragmenting source, $E_F$ its Fermi energy, and $A$ the mass of a produced fragment, which will enter the evaporation stage. This formula is incorporated in our model.

\subsubsection{Radial expansion}
The effect of the radial expansion is a lowering of the Fermi energy $E_F$ and of the Fermi momentum $p_F$ down to values $E_{\mathrm{Fbu}}$ and $p_{\mathrm{Fbu}}$ according to the following expressions: 
\begin{equation}
E_{\mathrm{Fbu}}=E_F \cdot \left( \frac{V_0}{V_{\mathrm{bu}}} \right)^{2/3}
\label{Ef-bu}
\end{equation}
\begin{equation}
p_{\mathrm{Fbu}}=p_F \cdot \left( \frac{V_0}{V_{\mathrm{bu}}} \right)^{1/3} \mbox{ .}
\label{pf-bu}
\end{equation}
In above equations, $V_0$ is the normal volume, while $V_{\mathrm{bu}}$ is the volume at freeze-out. The generally accepted volume increase at freeze-out is still not well established, and values between one and six times the normal volume are often considered~\cite{smm}.

\subsubsection{Coulomb expansion}
Furthermore, we shall consider the fact that the clusters formed in the thermal break-up are charged particles. Thus, Coulomb forces between these fragments have to be taken into account. 

K. C. Chung {\it et al.} developed ideas regarding the effects of this Coulomb expansion in nuclear fragmentation~\cite{coulomb}. We summarize here the main ideas and results of their article. 

Assuming that the expansion is uniform, but that the fragment radii do not change during expansion, it is possible to express the Coulomb contribution to the final kinetic energy: for a fragment of nuclear charge $Z$ and mass $A$, at a distance $r$ from the center of a prefragment of mass $A_{\mathrm{pre}}$, nuclear charge $Z_{\mathrm{pre}}$ and radius $R_{\mathrm{pre}}$, one has: 
\begin{equation}
E_{\mathrm{Coul}} = \frac{Z Z_{\mathrm{pre}} e^2}{R_{\mathrm{pre}}^3}r^2(1-A/A_{\mathrm{pre}})^2 \mbox{ .}
\label{coul}
\end{equation}
Equation \ref{coul} is remarkably simple, especially owing to the fact that the only relevant initial condition is the relative position of the fragment ($A$,$Z$) inside the system at break-up.  

While that kind of contribution can easily be included in a Monte-Carlo  simulation, its inclusion in an analytical model may be more complex, or require some assumptions that we will discuss in the next sections.

\subsection{Overview on the different fragmentation stages}
\begin{figure}[h]
\centering
\includegraphics*[width=65mm]{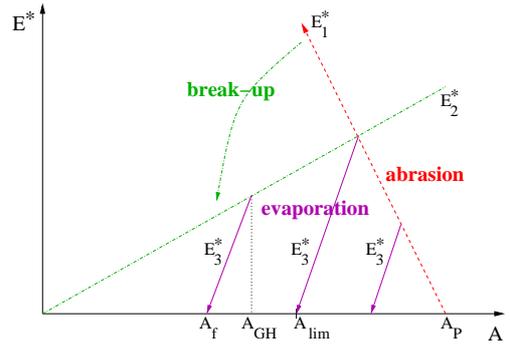}
\caption[]{\label{excitation}\footnotesize{Evolution of the excitation energy of the fragment through different reaction stages, from a given projectile of mass $A_p$. The abrasion process is represented by the line $E_1^*=\Delta E \cdot (A_p - A)$. If multifragmentation occurs, the system after freeze-out lands on the line $E_2^*=A/K\cdot T_{\mathrm{bu}}^2$. In all cases, evaporation goes on average along lines of the form $E_3^*=\varepsilon \cdot (A-A_f)$. See text for details. }}
\end{figure}

The impact parameter of the collision determines the mass and the excitation energy of the prefragment and hence the mass of the final fragment~\cite{hubele-zphysa}. 
Therefore, one can make the connection between the mass of the final residue and the conditions at the beginning of the evaporation stage~\cite{freeze}. This is schematically shown in fig. \ref{excitation}. 

The different stages of the reaction are connected with a reduction of mass. To follow the evolution of a prefragment, one has to read fig. \ref{excitation} from right to left. 
The first process of the fragmentation reaction, the abrasion stage, reduces the mass of the projectile, respectively target nucleus, inducing substantial excitation energy. The average excitation energy follows the line $E_1^*$, which corresponds to $\Delta E$=27~MeV induced by abrasion of one nucleon, as mentioned above. 

If the spectator after abrasion is left with an excitation energy of more than about 3 $A$~MeV, it breaks up into several fragments simultaneously. The products of this multifragmentation process are formed with a freeze-out temperature $T_{\mathrm{bu}}$ of about 5~MeV~\cite{freeze}. Before entering the evaporation process, these fragments fall on the freeze-out line $E_2^*$. 
The relation between temperature and excitation energy is described by the following formula: 
\begin{equation}
E^* = aT^2 \mbox{ .}
\label{temp}
\end{equation}
In this equation, $a$ is the level-density parameter; it is evaluated in a simplified way as $A/K$~MeV$^{-1}$. With a value of the inverse level-density parameter $K$=11~MeV and a freeze-out temperature of 5~MeV, we obtain $E^*_2=\frac{25}{11}A$~MeV. 

As soon as the system is excited (after abrasion or after multifragmentation) above the particle emission threshold, it has the possibility to emit neutrons, light charged particles and light nuclei by evaporation. This deexcitation process is represented by the arrows $E_3^*$. The link between the observed mass $A_f$ and the mass of the prefragment before evaporation is given by $E_3^*=\varepsilon \cdot (A-A_f)$, where $\varepsilon$ is the parameter representing the average energy consumed per evaporated nucleon. We assume a value of $\varepsilon$=15~MeV (see section~\ref{evaporation}). 

Figure \ref{excitation} suggests that final fragments lighter than a certain limiting mass $A_{\mathrm{lim}}$ are products of multifragmentation, while fragments heavier than this limit should be regarded as simple abrasion-evaporation residues. 

In other words, the present model considers two regions in mass, corresponding to the two regimes preceeding evaporation. They are divided by $A_{\mathrm{lim}}$ which is obtained  by the projection following an evaporation line $E_3^*$ of the intersection of lines $E_1^*$ and $E_2^*$. The solution of this geometrical requirement gives: 
%
\begin{equation}
A_{\mathrm{lim}}=\frac{K\cdot \varepsilon -T_{\mathrm{bu}}^2}{K\cdot \varepsilon}\cdot \frac{K\cdot \Delta E}{(K\cdot \Delta E + T_{\mathrm{bu}}^2)} \cdot A_p \mbox{ .}
\label{alim}
\end{equation}

Again, final residues with masses above $A_{\mathrm{lim}}$ are the products of a pure abrasion-evaporation process, while final residues with masses below this limit have experienced a multifragmentation process.

%% file: model.tex
\section{Analytical model}
\label{model}

Our goal is to obtain an analytical description of the momentum dispersion of fragmentation residues. Such an approach requires some approximations. 

We have seen that formula~\ref{goldhaber} represents only the abrasion stage of the fragmentation reaction and does not include the evaporation nor the multifragmentation stages. 

Nevertheless, formula~\ref{goldhaber} serves as a basis for the present work, since the Fermi motion is the dominant effect. 

Most of our results come from geometrical deductions of the reaction scheme depicted in fig.~\ref{excitation}. We already expressed the mass limit under which an observed fragment has presumably undergone break-up (eq.~\ref{alim}). 

We denote the mass of the prefragment just before evaporation by $A_{\mathrm{GH}}$ (GH stands for Goldhaber), since we will use it as a variable in Goldhaber's formula. In the abrasion-evaporation scenario, the mass before evaporation is obtained at the intersection between lines $E^*_1$ and $E^*_3$ (see fig.~\ref{excitation}). In the abrasion-multifragmentation-evaporation scenario, the mass before evaporation is the solution of the intersection of lines $E^*_2$ and $E^*_3$. 
For final-fragment masses close to the projectile ($A_f \geq A_{\mathrm{lim}}$), we obtain: 
\begin{equation}
A_{\mathrm{GH}}=\frac{\varepsilon A_f+\Delta E A_p}{(\Delta E+\varepsilon)} \mbox{ ,}
\label{pre}
\end{equation}

and if $A_f < A_{\mathrm{lim}}$, 
\begin{equation}
A_{\mathrm{GH}}=\frac{K\cdot \varepsilon}{K\cdot \varepsilon-T_{\mathrm{bu}}^2}\cdot A_f \mbox{ .}
\label{bu}
\end{equation}

In the abrasion stage, the influence of the Fermi motion is described by the Goldhaber formula (eq.~\ref{goldhaber}), and the standard deviation of the longitudinal momentum distribution of abrasion-evaporation products including the change in mass due to evaporation is given by the following formula:
\begin{equation}
\sigma^2_{p_{\parallel \mathrm{Fermi}}}={\left( \frac{A_f}{A_{\mathrm{GH}}} \right)}^2 \cdot \frac{p_F^2}{5} \cdot \frac{A_{\mathrm{GH}}(A_p-A_{\mathrm{GH}})}{(A_p-1)} \mbox{ .}
\label{abra}
\end{equation}

The Fermi momentum $p_F$ in the above equation is directly obtained from eq~\ref{fermi}. On the other hand, in the case of multifragmentation, one must consider that the Goldhaber formula (eq. \ref{goldhaber}) does not account for the volume expansion during the break-up nor for the finite temperature of the multifragmenting source. The volume expansion during the break-up is described via equations \ref{Ef-bu} and \ref{pf-bu}, where we assume a freeze-out volume equal 1.5 times the normal volume. The influence of finite temperature of the multifragmenting source is described according to Bauer's analytical model presented by eq.~\ref{temp_bauer}, where we assume a value of 5~MeV for the temperature at freeze-out $T_{\mathrm{bu}}$~\cite{freeze}. 

Since the descriptions of abrasion and break-up are both based on the Fermi momentum, we can adapt the expression of $p_{\mathrm{Fbu}}$, into an apparent Fermi momentum: 
\begin{equation}
\widetilde{p}_{\mathrm{Fbu}}=p_F \cdot \left( \frac{V_0}{V_{\mathrm{bu}}} \right)^{1/3} \cdot \left[ 1 + \frac{5\pi^2}{12} \left( \frac{T_{\mathrm{bu}}}{E_{\mathrm{Fbu}}} \right)^2\right] \mbox{.}
\label{pf-bu-bauer}
\end{equation}

The fact that we use an {\em apparent} Fermi momentum in case of multifragmentation requires that we write the corresponding contributions to the momentum dispersion in two terms: one contribution from the abrasion process (formation of a prefragment of mass $A_{\mathrm{bu}}$ out of a projectile $A_p$), and another contribution from the multifragmentation (formation of a product of mass $A_{\mathrm{GH}}$ out of an ensemble of $A_{\mathrm{bu}}$ nucleons). 

In case of multifragmentation, $A_{\mathrm{GH}}$ is expressed by eq.~\ref{bu}, but we do not know {\em a priori} the mass $A_{\mathrm{bu}}$ of the source of multifragmentation residues. As a reasonnable estimate, one can take the heaviest fragment that can enter multifragmentation. Let us recall that the abrasion stage is inducing an increase of excitation energy in the prefragment, by the amount of about $\Delta E=27$~MeV per abraded nucleon. Once the critical point leading to multifragmentation is reached, further abrasion of a few nucleons should lead to a break-up into smaller pieces, tending rapidly to vaporization. Final multifragmentation residues larger than alpha particles are then most probably produced by a prefragment of mass close to the theoretical maximum. 

We calculated the mass $A_{\mathrm{bu}}$ of the heaviest system undergoing multifragmentation. It corresponds to the collisions in which the abrasion has introduced enough energy to excite the nucleus to the critical temperature of 5 MeV, and is obtained from the intersection of the lines $E^*_1$ and $E^*_2$ shown in figure~\ref{excitation}:  
\begin{equation}
A_{\mathrm{bu}}=\frac{K\cdot \Delta E}{K\cdot \Delta E + T_{\mathrm{bu}}^2}A_p \mbox{ .}
\label{mother}
\end{equation}

Thus, the total contribution from the Fermi motion in the abrasion and the break-up on the kinematics of multifragmentation products is written: 
\begin{eqnarray}
\sigma^2_{p_{\parallel \mathrm{Fermi}}} & = & {\left( \frac{A_f}{A_{\mathrm{bu}}} \right)}^2 \!\! \cdot \frac{p_F^2}{5} \cdot \frac{A_{\mathrm{bu}}(A_p-A_{\mathrm{bu}})}{(A_p-1)} \nonumber \\
& + & {\left( \frac{A_f}{A_{\mathrm{GH}}} \right)}^2 \!\! \cdot \frac{\widetilde{p}_{\mathrm{Fbu}}^{\,2}}{5} \cdot \frac{A_{\mathrm{GH}}(A_{\mathrm{bu}}-A_{\mathrm{GH}})}{(A_{\mathrm{bu}}-1)}
\label{abra-bu}
\end{eqnarray}
with $\widetilde{p}_{\mathrm{Fbu}}$ the apparent Fermi momentum at break-up, given by eq.~\ref{pf-bu-bauer}.  
%

Apart from the effects of the Fermi motion, we mentioned that in case of multifragmentation, the expansion due to Coulomb forces should be implemented in the model. 
Let us reconsider eq.~\ref{coul}. 
Many variables are subject to fluctuations from one event to another, but through simple assumptions, one can simplify this equation. 
Isospin thermometer experiments~\cite{therm,freeze} have shown that the nuclear
composition of prefragments just before the sequential evaporation steps is
compatible with a scenario in which the prefragment keeps ``in memory'' the ratio $Z_p/A_p$ of the projectile. Both processes, abrasion and multifragmentation, preserve this value on average. 
It means that both the mother nucleus and its considered daughter have on average the same proportion of neutrons and protons. 
We express then the atomic number of these two nuclei in the following form:
\begin{equation}
Z=A \cdot \frac{Z_p}{A_p}
\label{isospin}
\end{equation}
where $A,Z$ design the features of each considered part of the nuclear ensemble. The effect of neutron distillation~\cite{neutron-distillation,isospin-distillation} on the mean $N/Z$ ratio of the fragments, which is small compared to the effect of evaporation, was neglected. 
Let us stress that this ratio is not preserved by the evaporation process, so that this formula does not apply to the final fragment. 

The system entering multifragmentation was taken to be the heaviest, with its mass $A_{\mathrm{bu}}$ given by eq.~\ref{mother}; its charge $Z_{\mathrm{bu}}$ is determined by the fixed ratio $Z/A$. 
This heaviest mother nucleus will generate the biggest electric field. This assumption should give us an upper estimate of the Coulomb repulsion, and at least allow us to check whether the magnitude of such an effect is compatible with the observed velocity distributions. 
%

The distance $r$ between the fragments at their formation is also crucial. We assume that the production of a fragment has a uniform probability inside the volume of the system. The variance of such a distribution of probability in one direction (in our case, along the longitudinal axis) is then the one of a uniform distribution between an interval [-$r_{\mathrm{max}}$;$+r_{\mathrm{max}}$]; such variance takes the form $(2 r_{\mathrm{max}})^2 / 12$. The integration of the electrical field in the spherical charged source up to the bounds of this interval leads to a Coulomb contribution to the longitudinal momentum that we name $P_{\mathrm{Coul}}$. 
The Coulomb contribution to the variance of the spectra is then written: 
\begin{equation}
\sigma^2_{p_{\parallel \mathrm{Coul}}}=\frac{1}{3} \cdot P_{\mathrm{Coul}}^2
\label{coulombmomentum}
\end{equation}
with the contribution to the momentum (see eq.~\ref{coul}) taken as: 
\begin{equation}
P_{\mathrm{Coul}}^2=2 A_{\mathrm{GH}} \cdot \mathrm{u} \cdot \frac{A_{\mathrm{GH}} A_{\mathrm{bu}}Z_p^2 e^2}{A_p^2 R_{\mathrm{bu}}^3}\cdot r_{\mathrm{max}}^2 \cdot (1-\frac{A_{\mathrm{GH}}}{A_{\mathrm{bu}}})^2
\label{full coulomb} 
\end{equation}
with $\mathrm{u}=931.5$ MeV/c$^2$ and $e^2=1.44$ MeV$\cdot$fm. $r_{\mathrm{max}}$ is the maximum distance between the centre of mass of the fragment and the one of the whole system, so it is the radius of the mother nucleus, reduced by the radius of the produced piece: $r_{\mathrm{max}}=R_{\mathrm{bu}}-R_{\mathrm{GH}}$. 

The further refinement in the description of the processes undergone by the observed fragment is related to the dynamics acting in the evaporation step. The sequential aspect of the evaporation gives rise to recoil momentum, broadening the momentum distributions. 

For this approach, we look first at the variance of the longitudinal velocity distributions. 
If we start from the prefragment, we may assume that the variance after evaporation of one nucleon is modified by the induced recoil momentum; the additional contribution to the variance of the velocity distribution is: 
\begin{equation} 
\sigma^2_{v_1}=\frac{p_{\mathrm{evap}}^2}{(A_{\mathrm{GH}}-1)^2} \mbox{ .}
\label{recoil1}
\end{equation}
We suppose that a single particle is evaporated with the mean momentum $p_{\mathrm{evap}}$. 
Then for a number $n$ of emitted particles, we can write: 
\begin{equation} 
\sigma^2_{v_n}=p_{\mathrm{evap}}^2 \cdot \sum_{i=0}^n\frac{1}{(A_{\mathrm{GH}}-i)^2} \mbox{ .}
\label{recoiln}
\end{equation}
For a large number $A_{\mathrm{GH}}$ (compared to the number of evaporated particles $n$), the difference between two steps of the sum is rather small. The sum can then be approximated by an integral. With the upper bound of the integral being $n=A_{\mathrm{GH}}-A_f$, we obtain this integrated relation for the final fragment: 
\begin{equation} 
\sigma^2_{v_{\mathrm{evap}}}=p_{\mathrm{evap}}^2 \cdot \left(\frac{1}{A_f}-\frac{1}{A_{\mathrm{GH}}}\right) \mbox{.}
\label{integrated}
\end{equation}
Since for the momentum variance, we have: 
\begin{equation}
\sigma^2_{p_{\mathrm{evap}}}={A_f}^2 \sigma^2_{v_{\mathrm{evap}}} \mbox{,}
\end{equation}
we can deduce the following, expressing $p_{\mathrm{evap}}^2$ along the longitudinal axis as in eq.~\ref{ratio} (with $\eta^2$ evaluated in eq.~\ref{eta}): 
\begin{equation} 
\sigma^2_{p_{\parallel \mathrm{evap}}}=A_f^2 \cdot \frac{p_F^2
\eta^2}{5} \cdot \left(\frac{1}{A_f}-\frac{1}{A_{\mathrm{GH}}}\right) \mbox{.}
\label{evapmomentum}
\end{equation}
$\sigma^2_{p_{\parallel \mathrm{evap}}}$ represents the contribution from the recoil of evaporated particles to the variance of the longitudinal momentum; $A_{\mathrm{GH}}$ has to be taken as in equations~\ref{pre} and~\ref{bu}.

Our final analytical formula is the quadratic sum of the different contributions, the Fermi motion (from eq.~\ref{abra} for masses beyond $A_{\mathrm{lim}}$ or eq.~\ref{abra-bu} for masses below $A_{\mathrm{lim}}$), Coulomb repulsion between the multifragmentation products (eq.~\ref{coulombmomentum}, applied below $A_{\mathrm{lim}}$), and the recoil induced by evaporation (formula~\ref{evapmomentum} for the whole mass range). 
The longitudinal momentum dispersion of final fragments is expressed by: 
\begin{equation}
\sigma^2_{p_\parallel} = \left\{ 
\begin{array}{l l}
\sigma^2_{p_{\parallel \mathrm{Fermi}}}+\sigma^2_{p_{\parallel \mathrm{evap}}} & \quad \mbox{beyond $A_{\mathrm{lim}}$}\\
 & \\
\sigma^2_{p_{\parallel \mathrm{Fermi}}}+\sigma^2_{p_{\parallel \mathrm{evap}}}+\sigma^2_{p_{\parallel \mathrm{Coul}}} & \quad \mbox{below $A_{\mathrm{lim}}$}\\ \end{array} \right.
\label{final}
\end{equation}
\begin{figure}[h]
\centering
\includegraphics*[width=85mm,height=130mm]{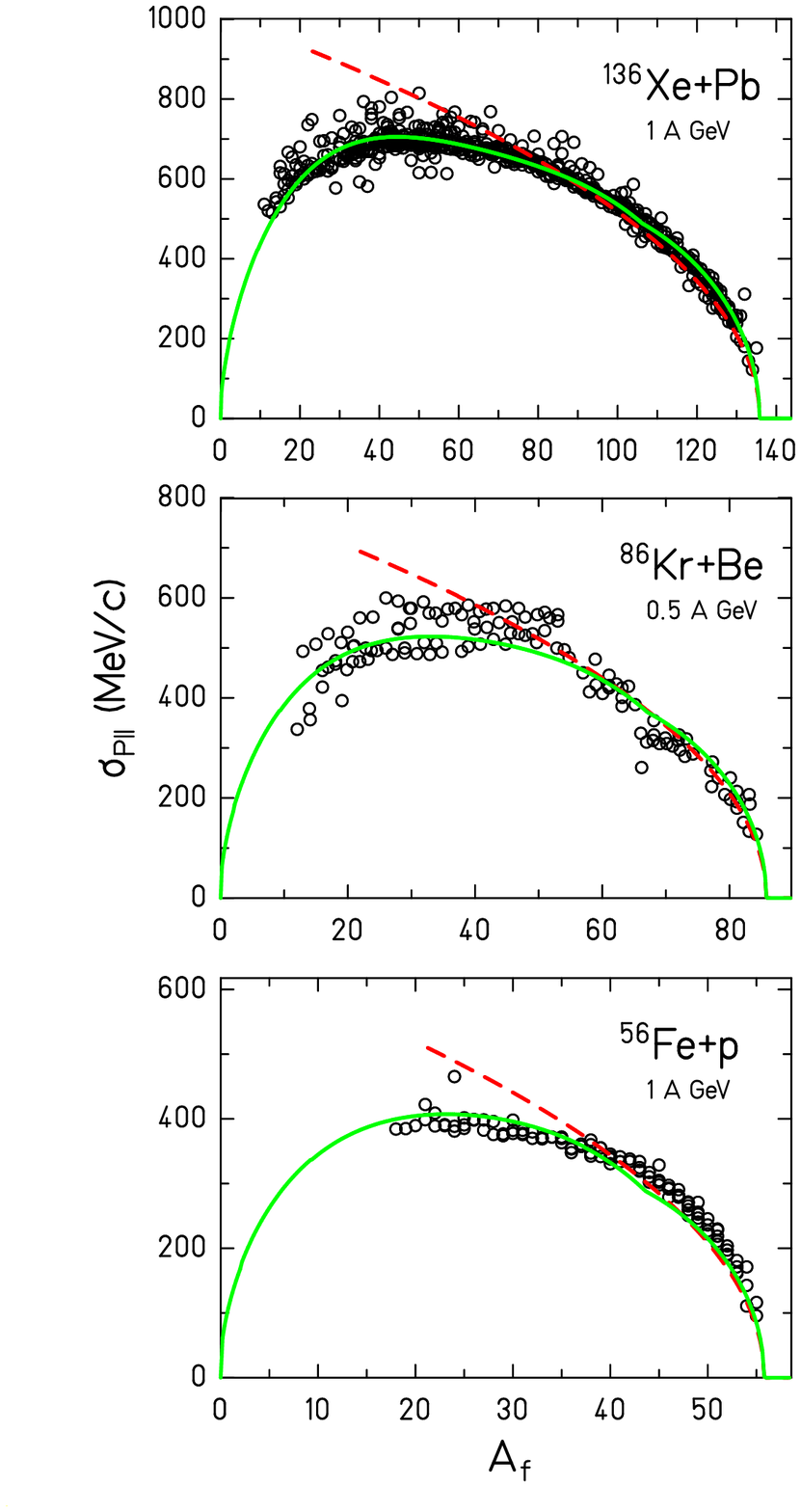}
\caption[]{\label{last}\footnotesize{Model in its full form as the square-root of eq.~\ref{final} (plain line) and Morrissey's systematics (dashed line) compared to different data sets~\cite{weber-exp,villagrasa,daniela-prc}. }}
\end{figure}

In fig.~\ref{last}, we confront our predictions with three sets of data, measured in the reactions $^{136}$Xe+Pb at 1~$A$~GeV~\cite{daniela-prc}, $^{86}$Kr+Be at 0.5~$A$~GeV~\cite{weber-exp} and $^{56}$Fe+p at 1~$A$~GeV~\cite{villagrasa}  respectively. These data present longitudinal momentum dispersion of fragmentation residues over a broad mass range. Other published data concentrating only on very peripheral collisions (i.e. masses very close to the mass of the projectile) are easily reproduced by our model as well as by Morrissey's systematics. Since the present model does not consider the fission process, data containing fission fragments were disregarded. 

The shape of the global prediction for $\sigma_{p_\parallel}$ of spectator
fragments is in good accordance with the overall trend of the data available for
this observable (see the confrontation with some data sets in fig.~\ref{last}). 


It is interesting to note that, although the abrasion picture, which is a basis of the present model, is not expected to be valid for targets as light as protons, our model is able to reproduce such data (see fig..~\ref{last}) with a remarkably good agreement. 

The degree of sophistication seems sufficient to reproduce the data. Our analytical formula is in general accordance with the data, even for light fragments and no systematic deviation is observed. 
In this mass region, where small residues can originate from more violent collisions, Coulomb interaction is reduced by the expansion of the system at break-up and our prediction could have been too large; this is not the case. 
%
%

The good agreement of our model with rather comprehensive sets of
experimental data give some confidence that the ingredients of our model
may correctly represent the physics of the fragmentation reaction. A
final conclusion, however, would require to investigate the sensitivity
of the results to the ingredients of the model and to the values of the
parameters. Such study is beyond the scope of the present work.

%% file: conclusion.tex
\section{Conclusion}
A new description of the width of the momentum distributions of fragmentation residues at relativistic energies is proposed. It describes the experimental data over a very broad mass range and includes most features of the known physical processes at play, constituting a great progress compared to previous formulae. 

The present new description has several advantages compared to predictions traditionally used. Morrissey provided an analytical description of the data but only in the limit of small mass losses. Furthermore, his formula is empirical and gives no insight into the physics of the reaction. 
Goldhaber introduced the idea that the Fermi motion provides the major contribution to the shape of the momentum distribution of the surviving fragments. The problem is that the quantity referred to in his model is not the final mass of the fragment which is observed. 

Going deeper into the description of the processes occurring in the collisions and using results provided by experiments, we suggested a new formula. It considers all relevant processes the system is undergoing in an analytical formulation. This description reproduces the data very well. 

Our formula gives a good estimate of the kinematics of projectile-like fragments in the relativistic-energy domain, which is the energy range concerning the applications mentioned in the introduction. 
Moreover, the physics basis of our description provides a realistic prediction of the longitudinal momentum width of fragmentation products, in the whole energy regime where the abrasion is a valid model for the collision phase, i.e. for energies above the Fermi energy.

%% file: appendix.tex
\clearpage
\onecolumn
\section*{Appendix}
To ease the use of the present model for practical applications, we propose in the present appendix the summary of needed equations, to calculate the momentum width of the fragmentation residues. We also list the needed parameters with their recommended values. 

The present model is valid over a large energy range (from about 50~A~MeV to at least 10~A~GeV), where fragmentation (and multifragmentation) processes occur. It applies to all kinds of projectiles (from light nuclei to uranium), independently from the target nucleus (therefore suited for any target, from protons to uranium). 
\\
\\
\textbf{Input parameters: }\\ 
$A_p$ : projectile mass \\
$Z_p$ : projectile atomic number \\
$A_f$ : final fragment mass \\
The predictions are calculated for  projectile-like fragments in the projectile frame, but if $A_p$ and $Z_p$ are replaced by the target mass $A_t$ and nuclear charge $Z_t$, the predictions apply to target-like fragments in the laboratory frame. \\
\\
\textbf{Calculated output: }\\
$\sigma^2_{p_\parallel}$ : dispersion of the linear momentum along the longitudinal axis for the final fragment (product of fragmentation reactions, excluding fission and fission-like processes). \\
\\
\textbf{Model parameters: }\\
$\mathrm{r_0}$ : radius = 1.4~fm\\
$e^2$ : squared elementary charge = 1.44~MeV$\cdot$fm\\
$\mathrm{u}$ : atomic mass unit = 931.5~MeV/$c^2$\\
$\Delta E$ : average induced excitation energy per abraded nucleon = 27~MeV\\
$<E_{\mathrm{fermion}}>$ : mean energy of one nucleon from Fermi motion = 20~MeV\\
$K$ : inverse level-density parameter $K=A/a=11$~MeV\\
$\varepsilon$ : average consumed energy per evaporated particle = 15~MeV\\
$E^{\mathrm{therm}}_{\mathrm{kin}}$ : average contribution of the thermal motion to the kinetic energy of evaporated neutrons or protons = 8~MeV\\
$T_{\mathrm{bu}}$ : break-up temperature = 5~MeV\\
$V_{\mathrm{exp}}=V_{\mathrm{bu}}/V_0$ : volume expansion factor = 1.5\\
\\
\textbf{Summary of equations: }\\
For a given projectile ($A_p$,$Z_p$), we calculate: 
\begin{displaymath}
p_F = 281 \cdot (1-A_p^{-0.568}) \: \mathrm{MeV}/c \mbox{ ,}
\end{displaymath}
\begin{displaymath}
E_F = \frac{p_F^{2}}{2 \mathrm{u}} \: \mathrm{MeV} \mbox{ ,}
\end{displaymath}
\begin{displaymath}
\eta^2 = \frac{1}{2}\cdot\left[ E^{\mathrm{therm}}_{\mathrm{kin}} + \left( E^{\mathrm{therm}}_{\mathrm{kin}}+ \frac{Z_p e^2}{\mathrm{r_0}(A_p^{1/3}+1)} \right) \right] / <E_{\mathrm{fermion}}> \mbox{. }
\end{displaymath}
The mass limit for multifragmentation (see text) can also be defined: 
\begin{displaymath}
A_{\mathrm{lim}}=\frac{K \cdot \varepsilon-T_{\mathrm{bu}}^2}{\varepsilon}\cdot \frac{\Delta E}{K\cdot \Delta E + T_{\mathrm{bu}}^2} \cdot A_p \mbox{ .}
\end{displaymath}
\textbf{For $A_f \geq A_{\mathrm{lim}}$} (final-fragment masses close to the projectile), multifragmentation did not occur. We need the following equations: 
\begin{displaymath}
A_{\mathrm{GH}}=\frac{\varepsilon A_f+ \Delta E \cdot A_p}{\Delta E + \varepsilon} \mbox{ ,}
\end{displaymath}
\begin{displaymath}
\sigma^2_{p_{\parallel \mathrm{Fermi}}}={\left( \frac{A_f}{A_{\mathrm{GH}}} \right)}^2 \cdot \frac{p_F^2}{5} \cdot
\frac{A_{\mathrm{GH}}(A_p-A_{\mathrm{GH}})}{A_p-1} \mbox{ ,}
\end{displaymath}
\begin{displaymath} 
\sigma^2_{p_{\parallel \mathrm{evap}}}=A_f^2 \cdot \frac{p_F^2
\eta^2}{5} \cdot \left(\frac{1}{A_f}-\frac{1}{A_{\mathrm{GH}}}\right) \mbox{.}
\end{displaymath}
The total momentum dispersion for this mass range is given by: 
\begin{displaymath}
\sigma^2_{p_\parallel} = \sigma^2_{p_{\parallel \mathrm{Fermi}}} + \sigma^2_{p_{\parallel \mathrm{evap}}}\mbox{.}
\end{displaymath}
\textbf{For $\mathbf{A_f < A_{\mathrm{lim}}}$}, multifragmentation occurs. The following ingredients are required: 
\begin{displaymath}
E_{\mathrm{Fbu}}=\frac{E_F}{V_{\mathrm{exp}}^{2/3}} \mbox{,}
\end{displaymath}
\begin{displaymath}
\widetilde{p}_{\mathrm{Fbu}}=\frac{p_F}{V_{\mathrm{exp}}^{1/3}} \cdot \left[ 1 + \frac{5\pi^2}{12} \left( \frac{T_{\mathrm{bu}}}{E_{\mathrm{Fbu}}} \right)^2\right] \mbox{,}
\end{displaymath}
\begin{displaymath}
A_{\mathrm{bu}}=\frac{K \cdot \Delta E}{K \cdot \Delta E + T_{\mathrm{bu}}^2}\cdot A_p \mbox{ ,}
\end{displaymath}
\begin{displaymath}
A_{\mathrm{GH}}=\frac{K\cdot \varepsilon}{K\cdot \varepsilon-T_{\mathrm{bu}}^2}\cdot A_f \mbox{ .}
\end{displaymath}
Then we can calculate: 
\begin{displaymath}
\sigma^2_{p_{\parallel \mathrm{Fermi}}} = {\left( \frac{A_f}{A_{\mathrm{bu}}} \right)}^2 \!\! \cdot \frac{p_F^2}{5} \cdot
\frac{A_{\mathrm{bu}}(A_p-A_{\mathrm{bu}})}{A_p-1} + {\left( \frac{A_f}{A_{\mathrm{GH}}} \right)}^2 \!\! \cdot \frac{\widetilde{p}_{\mathrm{Fbu}}^{\,2}}{5} \cdot \frac{A_{\mathrm{GH}}(A_{\mathrm{bu}}-A_{\mathrm{GH}})}{A_{\mathrm{bu}}-1} \mbox{ ,}
\end{displaymath}
And also: 
\begin{displaymath} 
\sigma^2_{p_{\parallel \mathrm{evap}}}=A_f^2 \cdot \frac{p_F^2
\eta^2}{5} \cdot \left(\frac{1}{A_f}-\frac{1}{A_{\mathrm{GH}}}\right) \mbox{.}
\end{displaymath}
%
Needed to calculate the contribution from Coulomb forces, the distance between the centre of mass of the produced fragment and the one of the whole system (i.e. the multifragmentation source) is taken to be 
\begin{displaymath}
r_{\mathrm{max}}=R_{\mathrm{bu}} - R_{\mathrm{GH}} \mbox{, }
\end{displaymath}
\begin{displaymath}
\mbox{with }R_{\mathrm{bu}}=\mathrm{r_0}\left(
A_{\mathrm{bu}}/V_{\mathrm{exp}} \right) ^{1/3} \mbox{ and }
R_{\mathrm{GH}}=\mathrm{r_0}\left( A_{\mathrm{GH}}/V_{\mathrm{exp}} \right) ^{1/3} \mbox{.}
\end{displaymath}
We define 
\begin{displaymath}
P^2_{\mathrm{Coul}}=2 A_{\mathrm{GH}} \cdot \mathrm{u} \cdot \frac{A_{\mathrm{GH}} A_{\mathrm{bu}}Z_p^2 e^2}{A_p^2
R_{\mathrm{bu}}^3}\cdot r_{\mathrm{max}}^2 \cdot \left( 1-\frac{A_{\mathrm{GH}}}{A_{\mathrm{bu}}}\right)^2 \mbox{,}
\end{displaymath}
entering the expression of the Coulomb contribution to the momentum dispersion: 
\begin{displaymath}
\sigma^2_{p_{\parallel \mathrm{Coul}}}=\frac{1}{3} \cdot P_{\mathrm{Coul}}^2 \mbox{. }
\end{displaymath}
The total momentum dispersion for this mass range is given by: 
\begin{displaymath}
\sigma^2_{p_\parallel} = \sigma^2_{p_{\parallel \mathrm{Fermi}}} + \sigma^2_{p_{\parallel \mathrm{evap}}} +
\sigma^2_{p_{\parallel \mathrm{Coul}}} \mbox{. }
\end{displaymath}

\twocolumn